\documentclass[showpacs,reprint,amsmath,amssymb,prb,floatfix]{revtex4-1}

\usepackage{graphicx}
\usepackage{dcolumn}
\usepackage{bm}
\usepackage{amsmath}

\begin{document}

\title{Color centers in NaCl by hybrid functionals} 
\author{Wei Chen}
\author{Christoph Tegenkamp}
\author{Herbert Pfn\"{u}r}
\email{pfnuer@fkp.uni-hannover.de}
\affiliation{Institut f\"{u}r Festk\"{o}rperphysik, Leibniz Universit\"{a}t Hannover, 30167 Hannover, Germany}
\author{Thomas Bredow}
\affiliation{Institut f\"{u}r Physikalische und Theoretische Chemie, Universit\"{a}t Bonn, 53115 Bonn, Germany}

\date{\today}

\begin{abstract}
We present in this work the electronic structure and transition energies (both thermodynamic and optical) of Cl vacancies in NaCl by hybrid density functionals.
The underestimated transition energies by the semi-local functional inherited from the band gap problem are recovered by the PBE0 hybrid functional through the non-local exact exchange, whose amount is adjusted to reproduce the experimental band gap.
The hybrid functional also gives a better account of the lattice relaxation for the defect systems arising from the reduced self-interaction.
On the other hand, the quantitative agreement with experimental vertical transition energy cannot be achieved with hybrid functionals due to the inaccurate descriptions of the ionization energies of the localized defect and the positions of the band edges.

\end{abstract}

\maketitle

\section{Introduction}

The anion vacancy is a prototypical point defect in alkali halide crystals, which can be introduced by heating them in an alkali metal vapor or by x-ray irradiation and by electron bombardment.
Anion vacancies in alkali halide are color centers ($F$ centers) because they are responsible for the coloration of the otherwise transparent crystals. \cite{Fowler1968}
The neutral anion monovacancy is the simplest form of the $F$ center, with a single bound electron in the vacancy center.
The electronic transition between the defect-induced levels can be trigged by the absorption of visible light.
For example, the presence of the absorption peak ($F$ band) associated with the $F$ center at 465 nm renders the NaCl crystal yellow. \cite{Pohl1937,Pick1938,Pick1958,Kittle2005}
An eminent application of anion vacancies in alkali halides due to the coloration is the color center laser first demonstrated in 1965 by Fritz and Menke. \cite{Fritz1965,Hecht1999}

Due to their exotic features and long history, $F$ centers in alkali halides have been studied extensively using various techniques such as optical absorption \cite{Rabin1960,Buchenauer1968,Perregaux1974}, 
Raman spectroscopy \cite{Worlock1965,Benedek1967,Ghomi1979,Leblans1987}, 
the Stark effect \cite{Grassano1970,Bogan1970,Stiles1970}, 
and luminescence \cite{Takiyama1978,Fujita1980,Koyama2009}.
It is now well established that the defect states induced by anion vacancies in alkali halides are localized deep levels because of the small dielectric constant, large band gap, and high effective mass of the host crystals.
These localized defect states cannot be correctly described within the effective mass approximation. \cite{Yu2005}
Early calculations employed either a linear combination of atomic orbital (LCAO) or the vacancy-centered variational wavefunction method within the Hartree-Fock (HF) theory.
The $F$ center was treated as a hydrogenic system trapped in the Coulomb potential created by the ions as point charges. \cite{Gourary1957,Blumberg1958,Wood1961,Wood1964}
Although these calculations agree well with experimental data, it was later pointed out by Murrel and Tennyson, on basis of their LCAO self-consistent field (SCF) calculations within embedded cluster method, that the agreement could be fortuitous, provided the results are sensitive to the selection of model Coulomb potential surrounding the anion vacancy. \cite{Murrel1980}
Thus it is desirable to use a "true" potential from \textit{ab initio} calculations.
Numerous LCAO SCF cluster calculations have been performed at the restricted open-shell (ROHF) \cite{Chaney1976,Murrel1980} and unrestricted HF (UHF) level. \cite{Kung1982,Pandey1990,Leitao2002}
The improvements, however, are limited and not unanimous as a result of the lack of electronic correlations and of the small cluster size limited by the computational capability.

Recent advances in density functional theory (DFT) makes it the tool of choice for studying the electronic structure of point defects in insulators.
One major deficiency of DFT is that the standard local and semi-local approximations for the exchange-correlation (XC) functional suffer from the spurious electron self-interaction and the lack of the derivative discontinuities of the XC potential with respect to the particle number \cite{Kummel2008}.
The calculated band gap is therefore generally smaller than experimental value, and the underestimation is more pronounced for wide band gap insulators.
For NaCl, DFT calculations predict band gaps from 4.5 eV to 5.0 eV \cite{Li2007,Fang2008,Chen2009}, much smaller than the experimental value (8.5 eV) \cite{Poole1975}.
The band-gap problem poses serious uncertainties to the positions of the defect electronic levels and makes the credibility of the calculation questionable.
One of the simplest yet the crudest way to overcome the gap problem is to apply a scissors operation to the conduction band and shift it rigidly to the experimental value relative to the valence band maximum (VBM).
However, this still leaves ambiguities to the position of defect levels. 
In particular, for deep levels which do not follow either the characteristics of the VBM or conduction band minimum (CBM), the scissors scheme is not well-justified and not satisfactory in first-principles calculations.
Other physically intuitive approaches to the problem include self-interaction corrections (SICs) \cite{Perdew1981}, and the DFT$+U$ method where the localized states with a strong Coulomb repulsion, such as incomplete $d$ or $f$ shells in transition-metal oxides, are modeled by a Hubbard-like $U$ term \cite{Liechtenstein1995} to open the band gap.

An alternative, and a more general approach to the gap problem is to relieve the unphysical self-interaction and derivative discontinuity problems by incorporating the non-local exact exchange (or Fock exchange) into the DFT XC functional.
One of the implementation of the exact exchange is the screened exchange (sX) method, which replaces the whole local-density approximation (LDA) exchange by a Thomas-Fermi screened Coulomb exchange potential \cite{Bylander1990}.
In hybrid density functionals, a fraction of exact exchange is mixed with LDA or generalized gradient approximation (GGA) exchange.
The hybrid functionals generally not only improve the band gap, but also yield better results in bulk properties such as lattice parameters, bulk moduli and heats of formation for semiconductors and insulators \cite{Paier2006,Paier2006a}. 
A recent study comments that hybrid functionals are superior to SICs in reducing self-interaction errors since the unitary invariance is preserved. \cite{Stengel2008}
Successful applications of hybrid functionals in the defect properties of various oxides and semiconductors have been recently reported. \cite{Saal2009,Stroppa2009,Gali2009,Agoston2009,Scanlon2009,Janotti2010,Deak2010}

In this work, we revisit the anion (Cl) vacancy in NaCl in different charge states (-1, 0 and +1) by both semi-local and hybrid functionals, aiming for a better understanding of the $F$ center in alkali halide and the performance of the hybrid functionals for localized defects.
We note that besides the band-gap problem, the finite-size effect arises within the current modeling scheme for the vacancy.
The common procedure for defect energetics calculation is to embed the defect into a supercell under periodic boundary condition (PBC).
The advantage of using supercells instead of cluster methods is that the band structure of the host crystal is well-defined, as the cell is bulk-like. \cite{Walle2004}
However, tractable DFT calculations are usually constrained to about 1000 atoms, 
and the size of system is further limited for hybrid functionals in a plane-wave basis set.
A single Cl vacancy in a 1000-atom NaCl supercell corresponds to a vacancy concentration of the order of $10^{20}$cm$^{-3}$, which is much higher than those found in experiment ($10^{15}$ to $10^{19}$ cm$^{-3}$).
The periodic images of the point defects in a high density thus give rise to unrealistic defect-defect interactions, making the defect energetics dependent on the size of the supercells.
The problem is even more serious for charged defects, as the neutralizing background slows the convergence of defect energies with respect to the supercell size. \cite{Makov1995}
Other sources of error for small supercells involve the elastic energy due to artificial relaxations of ions, and defect level dispersions introduced by defect-defect interactions.
Corrections for the finite-size effect have been found indispensable in defect calculations for realistic interpretations \cite{Lany2008,Nieminen2009,Lany2009,Castleton2009}, and they will be discussed and applied to the present study of Cl vacancies.

\section{Choice of hybrid functionals}

We first assess the performance of the hybrid functionals on the bulk properties of NaCl, and compare it to the GGA-PBE (Perdew-Burke-Ernzerhof) functional \cite{Perdew1996}.
Both unscreened and screened hybrid functionals are employed.
In the unscreened PBE0 functional, the exchange part of the XC energy $E_{xc}$ is constructed by mixing a fraction ($\alpha$) of non-local exact exchange $E_x$ with PBE exchange $E_x^\text{PBE}$, while the correlation energy is simply taken from PBE, $E_c^\text{PBE}$: 
\begin{equation}
E_{xc}^\text{PBE0}=\alpha E_x + (1-\alpha) E_x^\text{PBE} + E_c^\text{PBE}.
\end{equation}
The amount of exact exchange $\alpha$ is a variable from 0 to 1, although conventionally $\alpha=0.25$ is used as suggested by perturbation theory \cite{Perdew1996a}.
In practice $\alpha$ is usually varied to meet the experimental gap value.
In a plane-wave basis set, the evaluation of the exact (HF) exchange is a hog to the computational resources and tends to be rather slow because of its truly non-local nature.
The calculation can be accelerated by truncating the slowly decaying long-range part of the exact exchange as in the HSE (Heyd-Scuseria-Ernzerhof) hybrid functional \cite{Heyd2003}:
\begin{eqnarray}
\label{eq:hse}
\nonumber
E_{xc}^\text{HSE} &=& \alpha E_x^\text{sr}(\mu) + (1-\alpha) E_x^\text{PBE,sr}(\mu) \\* 
&& + E_x^\text{PBE,lr}(\mu) + E_c^\text{PBE}.
\end{eqnarray}
The screening parameter $\mu$ in Eq.~\eqref{eq:hse} determines the separation of the short-range (sr) and long-range (lr) parts:
\begin{equation}
\frac{1}{r} = \text{sr}(r)+\text{lr}(r) = \frac{1-\text{erf}(\mu r)}{r} + \frac{\text{erf}(\mu r)}{r}.
\end{equation}
In one limit when $\mu=0$, the long-range term is zero and HSE reduces to the unscreened PBE0 functional.
For $\mu \rightarrow \infty$, HSE is identical to GGA-PBE since the whole exact exchange is screened.
Here we use the optimized $\mu=0.207$ \AA$^{-1}$, following Ref.~\onlinecite{Krukau2006} along with $\alpha = 0.25$ and refer to this functional as HSE06.

\begin{table}
\caption{\label{bulk} Calculated lattice constant ($a_0$), fundamental band gap at $\Gamma$ ($E_g$), dielectric constant ($\epsilon_\infty$) and enthalpy of formation ($\Delta H_f$) of rocksalt NaCl using various DFT functionals.
}
\begin{ruledtabular}
\begin{tabular}{ldddd}
 & \multicolumn{1}{c}{$a_0 \text{(\AA)}$} & \multicolumn{1}{c}{$E_g \text{(eV)}$} & \multicolumn{1}{c}{$\epsilon_\infty$} & \multicolumn{1}{c}{$\Delta H_f \text{(eV)}$} \\
\hline
GGA-PBE & 5.69 & 5.00 & 2.33 & -3.69 \\
HSE06 & 5.65 & 6.43 & 2.13 & -3.85 \\
PBE0 $(\alpha=0.25)$ & 5.64 & 7.19 & 1.98 & -3.85\\
PBE0 $(\alpha=0.40)$ & 5.62 & 8.47 & 1.86 & -3.93\\
Expt. & 5.57\footnotemark[1] & 8.5\footnotemark[2] & 2.3\footnotemark[3] & -4.26\footnotemark[1]\\
\end{tabular}
\end{ruledtabular}
\footnotetext[1]{Reference \onlinecite{Paier2007}.}
\footnotetext[2]{Reference \onlinecite{Poole1975}.}
\footnotetext[3]{Reference \onlinecite{Bechstedt1988}.}
\end{table}

In Table~\ref{bulk} selected bulk properties of NaCl calculated using the GGA-PBE and hybrid functionals are summarized together with the experimental values.  
The calculations are carried out in the projector augmented wave (PAW) framework with the \textsc{vasp} code \cite{Kresse1996,Kresse1999,Paier2005}.
A semicore pseudopotential of Na is used, treating the $2p3s$ electrons as valence electrons.
The kinetic cutoff energy for the plane-wave basis set is 500 eV.
A $\Gamma$-centered $8\times8\times8$ Monkhorst-Pack \textbf{k}-point mesh \cite{Monkhorst1976} is applied to the primitive cell containing one formula unit of NaCl.
For HSE06 and PBE0 calculations, a down-sampled $4\times4\times4$ mesh is used to evaluate the non-local exact exchange.
The down-sampling for the non-local exchange reduces the computing time significantly. 
It is generally necessary for the PBE0 to have a finer \textbf{k}-point mesh than for the screened HSE functional to reach convergence.\cite{Paier2006,Paier2006a}
For the present case, the HF exchange using the PBE0 changes by roughly 15 meV per atom from the down-sampled $4\times4\times4$ to the full $8\times8\times8$, while the energy is already converged within $10^{-2}$ meV with the HSE06 functional.
Nevertheless, the choice of the down-sampled \textbf{k}-point for the PBE0 calculations is sufficient for the bulk properties. 
The lattice constant $a$ is determined when the residual force is smaller than 5 meV/\AA.
The high frequency macroscopic dielectric constant $\epsilon_\infty$ can be calculated within a $GW$ scheme using the random-phase approximation (RPA) with local field effect included.
Around 90 empty bands are used for calculating the dielectric constant.
The dielectric constant will also be referred to later for the finite-size corrections.
Finally, the formation energy $\Delta H_f$ is obtained
\begin{equation}
\label{eq:Hf}
\Delta H_f = E_\text{NaCl(s)} - E_\text{Na(s)} - \frac{1}{2} E_\text{Cl$_2$(g)}.
\end{equation}
In Eq.~\eqref{eq:Hf}, $E_\text{NaCl(s)}$ is the total energy of bulk NaCl.
$E_\text{Na(s)}$ is the energy of bulk Na in a body centered cubic (bcc), which was optimized and calculated using the same \textbf{k}-point mesh and cutoff energy as the bulk NaCl.
$E_\text{Cl$_2$(g)}$ refers to the energy of one gas phase Cl$_2$ molecule in a large tetragonal cell.

One immediately observes that the hybrid functionals improve not only the direct band gap ($\Gamma_{15} \rightarrow \Gamma_{1}$) but also the lattice constant and heat of formation compared to the GGA-PBE calculation in Table~\ref{bulk}, in agreement with earlier calculations.\cite{Paier2006,Paier2006a}
Yet, it is found the band gaps are still underestimated for the hybrid functionals with the original fraction (0.25) of exact exchange, and the PBE0 yields a much closer value to experiment than the HSE06.
This implies that for wide gap insulators, as the electronic screening is quite weak, the unscreened exact exchange in PBE0 is preferred.
For defect calculations, it is customary to tune the fraction of the exact exchange so that the experimental band gap can be reproduced.\cite{Oba2008,Janotti2010}
By increasing the amount of the non-local exchange from 0.25 to 0.40 within the PBE0, the band gap of NaCl recovers nearly to the experimental value, and the lattice constant and heat of formation are also reproduced best among the chosen functionals.
We note that hybrid functionals tend to underestimate the dielectric constant of NaCl, a trend also found for semiconductors and other insulators.\cite{Fuchs2007}
An accurate description of the electronic dielectric constant with the hybrid functionals will require an explicit account of excitonic effects.\cite{Shishkin2007,Paier2008}

To this end, we face several functionals for the subsequent calculations of the Cl vacancy in NaCl.
The PBE0 ($\alpha=0.40$) (we will refer it to mPBE0 hereafter) is apparently favored since it reproduces the experimental gap.
However, as the choice of the fraction of the non-local exact exchange is empirical to some extent, its impact on the position of the deep defect level for wide gap insulators is still unknown.
Meanwhile, the screened hybrid functional is of great interest as it shows considerable success in the prediction of defect energetics.
Therefore it is plausible to also include the HSE06 functional with the original $\alpha$, as well as the GGA-PBE for the defect calculations.

\section{\label{sec:ks}Electronic structure of Chlorine vacancies}

In this section we briefly sketch out the single-particle Kohn-Sham (KS) eigenvalues of the Cl vacancy induced electronic levels.
Supercells containing 64 atoms are employed for the calculations.
The Brillouin zone is sampled with a $2\times2\times2$ \textbf{k}-point mesh, and a plane-wave basis set cutoff energy of 450 eV is used.
Further, the \textbf{k}-point mesh for the non-local exact exchange is down-sampled to the $\Gamma$-point for HSE06, while a full $2\times2\times2$ \textbf{k}-point mesh is necessary for well converged energies in PBE0 calculations.
The convergence criterion for full relaxations is 0.01 eV/\AA.

\begin{figure}
\includegraphics{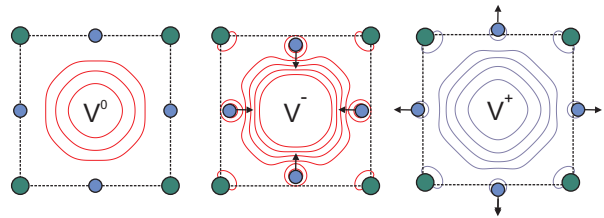}
\caption{\label{fig:chg}(color online). The electron density ($V^0$ and $V^-$) and hole density ($V^+$) isosurface of the $a_{1g}$ state in the (100) plane calculated with the mPBE0 functional.
The lines are drawn in intervals of 0.01$e$/\AA$^3$.
The displacements of the nearest neighbor atoms around the Cl vacancy after relaxation are illustrated by the arrows.
The blue and green circles represent the Na and Cl atoms, respectively.}
\end{figure}

The removal of one Cl atom in a perfect NaCl crystal leaves a neutral vacancy $V^0$ with one electron bound to the vacancy center.
The localized nature of the unpaired electron can be clearly identified in the charge density isosurface shown in Fig.~\ref{fig:chg}. 
The $1s$ characteristics of the wavefunction in the vacancy is contributed equally from the six neighboring Na atoms.
The negligible displacement of the neighboring atoms around $V^0$ (see Table~\ref{disp}) keeps the singly occupied $a_{1g}$ level unshifted after relaxation.
In the +1 charge state $V^+$, the $a_{1g}$ state is unoccupied and the polaronic hole is trapped in the vacancy.
The nearest neighbor Na atoms tend to relax away from the vacancy because of the positive electrostatic potential inside the vacancy.
The outward relaxation delocalizes the $a_{1g}$ (see Fig.~\ref{fig:chg}), and shifts it to higher energy towards the CBM.
In the -1 charge state $V^-$, the $a_{1g}$ state becomes doubly-occupied.
Upon relaxation the six nearest Na atoms show inward displacement towards the vacancy (Fig.~\ref{fig:chg}) as a polaronic distortion.
As a consequence, the two electrons are more localized inside the vacancy site, shifting the $a_{1g}$ state to lower energy. 
We note that the relaxations of the neighboring atoms around the anion vacancy follow the $O_h$ symmetry for all charge states.
No symmetry lowering (or Jahn-Teller distortion) is found since the defect level is either singly occupied for $V^0$, or doubly occupied (unoccupied) for $V^-$ ($V^+$). 
The reduced self-interaction in hybrid functionals also results in more pronounced atomic displacements for the charged defects as seen in Table~\ref{disp}.

\begin{table}
\caption{\label{disp} Displacements (in the equilibrium bond length 0.5$a_0$) of the nearest-neighbor Na atoms around the Cl vacancy calculated using the 64-atom cell. The positive value represents an outward relaxation against the vacancy, and vice versa. The values obtained with the 216-atom cell using the GGA-PBE are shown in parentheses.}
\begin{ruledtabular}
\begin{tabular}{lccc}
 & GGA-PBE & HSE06 & mPBE0 \\
\hline
$V^+$ & $+0.035$ $(+0.043)$ & $+0.035$ & $+0.036$ \\
$V^0$ & $-0.000$ $(-0.000)$ & $-0.000$ & $-0.000$ \\
$V^-$ & $-0.039$ $(-0.040)$ & $-0.041$ & $-0.044$ \\
\end{tabular}
\end{ruledtabular}
\end{table}

\begin{table}
\caption{\label{ks} Energy levels (in eV) of the single-particle Kohn-Sham $a_{1g}$ state of the Cl vacancy in a cubic 64-atom cell referenced to the VBM.
The energy is averaged over the BZ.
For $V^0$, the positions of the $a_1$ state in spin-up (occupied) and spin-down (unoccupied) channels are given.
The absolute positions of the host band edges are also given. }
\begin{ruledtabular}
\begin{tabular}{lcccccc}
& \multicolumn{2}{c}{PBE} & \multicolumn{2}{c}{HSE06} & \multicolumn{2}{c}{mPBE0} \\
\cline{2-3}
\cline{4-5}
\cline{6-7}
& rigid & relaxed & rigid & relaxed & rigid & relaxed \\
\hline
$V^+(a_{1g}^0)$             & 3.74 & 5.02 & 4.90 & 6.25 & 6.81 & 8.25 \\
$V^0(a_{1g}^1)_\uparrow $   & 4.09 & 4.10 & 4.83 & 4.83 & 5.30 & 5.30 \\
$V^0(a_{1g}^1)_\downarrow $ & 4.85 & 4.86 & 6.13 & 6.13 & 8.12 & 8.12\\
$V^-(a_{1g}^2)$             & 5.04 & 4.37 & 6.07 & 5.22 & 6.87 & 5.69 \\
$\epsilon_\text{VBM}$ & \multicolumn{2}{d}{-0.76} & \multicolumn{2}{d}{-1.63} & \multicolumn{2}{d}{-2.78} \\
$\epsilon_\text{CBM}$ & \multicolumn{2}{d}{4.24} & \multicolumn{2}{d}{4.80} & \multicolumn{2}{d}{5.69} \\
\end{tabular}
\end{ruledtabular}
\end{table}

Table~\ref{ks} summarizes the KS energies of the Cl vacancies in various charge states for both rigid and relaxed defect structures.
The choice of \textbf{k}-point gives rise to a dispersion of the electronic level within the finite-size supercell scheme.
The dispersion introduces a strong dependence on the supercell size of the energy level at the $\Gamma$-point $\epsilon^\Gamma$. 
Thus we average the defect level energy over the Brillouin zone $\overline{\epsilon}$ since the averaged level shows a much better convergence than $\epsilon^\Gamma$. \cite{Lany2008} 
The dispersion also slightly pushes the host CBM to higher energies.
The finite-size effect will be discussed in detail in Sec.~\ref{sec:corr}.
As predicted by all functionals, the $a_{1g}$ states in all charged states lie within the upper half of the host band gap.
We note that the absolute KS energies of the $a_{1g}$ states [$\epsilon_\text{KS}(a_{1g})+\epsilon_\text{VBM}$] are roughly unaffected when going from semi-local to hybrid functionals. 
In hybrid functionals, the VBM is lowered by 0.9 eV (HSE06) and 2.0 eV (mPBE0) with respect to the GGA-PBE as a result of the reduced self-interaction for the Cl $3p$ states.
On the other hand, for rigid structures, the widening of the band gap in hybrid functionals tends to place the $a_{1g}$ of $V^-$ further away from the CBM compared to the semi-local functional, while the unoccupied $a_{1g}$ state of $V^+$ is closer to the CBM when hybrid functionals are used.

\section{Thermodynamic transition energies and finite-size corrections}

In general the single-particle energy level of the defect as calculated from the KS equation differs from the experimentally observed transition energies. \cite{Lany2008}
A rigorous approach to the transition energies, as discussed in this section, relies on the total energy difference of the defect energetics in various charged states.

\subsection{Formalisms of formation energies and transition energies}

A central quantity for the defect energetics is the formation energy $E_f$ for a defect $D$ in charge state $q$ 
\begin{equation} 
\label{eq:Ef}
E_f = E_\text{D} - E_\text{H} + \Sigma n_i \mu_i + q(\epsilon_\text{VBM}+ \epsilon_\text{F}),
\end{equation}
where $E_\text{D}$ and $E_\text{H}$ are the total energy of the supercell with the defect $D$, and the host supercell without defects, respectively.
$n_i$ is the number of atoms removed from the supercell ($n_i =1$ for the Cl monovacancy) or the number of impurities added ($n_i < 0$).
$\mu_i$ refers to the chemical potential of the associated defect particle reservoir, and is subject to equilibrium conditions.
For the present study, under extreme Cl-rich (or equivalently Na-poor) conditions
\begin{equation}
\mu_\text{Cl}= \frac{1}{2} E_\text{Cl$_2$(g)}.
\end{equation}
This places an upper limit on $\mu_\text{Cl}$.
The lower bound can be deduced from the following relation:
\begin{equation}
\mu_\text{Na} + \mu_\text{Cl} = E_\text{NaCl(s)}.
\end{equation}
Therefore under Cl-poor (or Na-rich) conditions, which facilitate the formation of Cl vacancies, 
\begin{equation}
\mu_\text{Cl} \geq E_\text{NaCl(s)} - \mu_\text{Na(s)} = \Delta H_f + \frac{1}{2} E_\text{Cl$_2$(g)},
\end{equation}
and this sets the lower limit of $\mu_\text{Cl}$.

The remaining term $\epsilon_\text{VBM}+\epsilon_\text{F}$ in the formation energy [Eq.~\eqref{eq:Ef}] represents the chemical potential, or Fermi energy of the electrons in charged defects.
The Fermi energy $\epsilon_\text{F}$ is varied within the band gap referenced to the energy of the host VBM $\epsilon_\text{VBM}$ ($0 \leq \epsilon_\text{F} \leq E_g$).
Here $\epsilon_\text{VBM}$ is evaluated as the energy difference between a perfect host supercell and the same host supercell with one electron removed from the VBM:
\begin{equation}
\epsilon_\text{VBM} = E_\text{H}^{0}(n) -E_\text{H}^{+}(n-1),
\end{equation}
where $n$ is the number of electron in the host supercell.
In principle, one needs a sufficiently large supercell with $n \rightarrow \infty$ corresponding to the dilute limit.
In practice, a fractional charge $q$ can be used along with a small supercell to obtain the $\epsilon_\text{VBM}$
\begin{equation}
\epsilon_\text{VBM} = \lim_{q\to0}\frac{E_\text{H}^{0} -E_\text{H}^{q}}{q}.
\end{equation}
In the present case, the $\epsilon_\text{VBM}$ converges well within a 64-atom cell and a fraction charge of 0.001 $e$.

For charged defects, it is evident from Eq.~\eqref{eq:Ef} that the formation energy is dependent on the chemical potential of the exchanged electron.
The thermodynamic transition energy $\epsilon (q/q')$ is defined as the Fermi energy at which the charge state $q$ and $q'$ of the defect system can be transformed spontaneously from one to the other. 
Therefore at the transition energy $\epsilon (q/q')$ these two charge states have the same formation energy.
This gives the following form of the transition energy
\begin{equation}
\label{eq:trans}
\epsilon (q/q') = \frac{E_\text{D}(q)-E_\text{D}(q')}{q'-q} - \epsilon_\text{VBM}.
\end{equation}

\subsection{\label{sec:corr}Finite-size corrections}
Before proceeding to the results, we shall discuss the correction methods for the finite-size effect, whose causes have been already identified in Introduction.
For charged defects, the simplest correction is to align the electrostatic potential in the defect supercell to that of the host supercell.
This is usually done by inspecting the potential difference $\Delta V$ between the core potentials of the atoms far from the defect center and that of the bulk cell, and the energy correction term is essentially $\Delta E=q \Delta V$.
This correction is rationalized by the fact that in periodic supercell calculations the zero of the electrostatic potential is chosen arbitrarily for each calculation, and the charged defect gives rise to a constant shift in the potential so that the bulk VBM cannot be applied directly to the defect supercell.  
However, due to the small size of the supercell used even the atoms farthest from the charged defect center are not bulk-like, making such correction scheme inaccurate.
Recent study reveals that the potential alignment resembles the Makov-Payne scheme, \cite{Makov1995} whereas the latter targets the correction of the unphysical defect-defect interactions.
Indeed, Komsa and Rantala found that $\Delta V$ has the form $(\epsilon L)^{-1}$, where $L$ is the lattice constant of a cubic supercell. \cite{Komsa2009}
This is analogous to the Makov-Payne scheme to first order.

The popular Makov-Payne scheme for a charged defect in a cubic supercell in the dilute limit ($L \rightarrow \infty$) is expanded as
\begin{equation}
E_f (L) = E_f(L \rightarrow \infty) - \frac{\alpha_\text{Md} q^2}{2\epsilon L} - \frac{2 \pi q Q}{3\epsilon L^3} + O(L^{-5}),
\end{equation}
where $\alpha_\text{Md}$ is the Madelung constant dependent on the lattice type, and $q$ and $Q$ the monopole and quadrupole moment of the defect charge, respectively \cite{Makov1995}.
The first order term, also called the Madelung energy, is thus the correction to the monopole-monopole interaction arising from the periodic image.
The $L^{-1}$ behavior of the artificial electrostatic interaction vanishes slowly, and this is usually the leading source of error.
The higher order corrections have much smaller effects on the formation energy for ionic crystals, and it is usually accurate enough to include the corrections up to the quadrupole term.
We note that in the Makov-Payne scheme the defect states are assumed to be localized, which is the case for the Cl vacancies in NaCl.  
For delocalized levels higher order corrections might become necessary.
Although the Makov-Payne expansion is sound and accurate, it has been found that direct corrections using the Madelung energy and multipole interactions are prone to overshoot the formation energy, in particular for small supercells \cite{Castleton2006}.
A more reliable approach is to employ a scaling method by performing a series of calculations using supercells of different sizes with the same symmetry. \cite{Castleton2004}
The corrected formation energy $E_f (L \rightarrow \infty)$ then can be extrapolated to the dilute limit by fitting the calculated formation energies within finite-size cells to
\begin{equation}
E_f (L) = E_f(L \rightarrow \infty) + a_1 L^{-1} + a_3 L^{-3},
\end{equation}
where $a_n$ and $E_f(L \rightarrow \infty)$ are fitting parameters.
It is clear that this scaling law method requires at least 4 supercells, and is rather computationally laborious.

In a recent work Freysoldt \textit{et al.} proposed a general correction scheme (we will refer to it as the FNV scheme hereafter) for finite-size effect based on a single calculation of defect supercell without empirical parameters \cite{Freysoldt2009}:
\begin{equation}
E_f = E_f(L \rightarrow \infty) + E_q^\text{latt}-q \Delta_{q/b},
\end{equation}
where $E_q^\text{latt}$ is the macroscopically screened lattice energy of the defect charge $q_d$ with compensating background, and $\Delta_{q/b}$ is an alignment term referenced to the bulk supercell to account for the microscopic screening.
As the long-range $E_q^\text{latt}$ scales as $L^{-1}$ and the short-range alignment term as $L^{-3}$, the FNV scheme can be seen as an extension to the Makov-Payne expansion.
It also allows for an explicit expression for the third-order $L^{-3}$ energy term.

Now we apply both Makov-Payne scaling and FNV schemes to the formation energies of the charged Cl vacancies ($V^+$ and $V^-$) in NaCl.
We refrain from including the potential alignment in these two schemes in order to avoid double-counting of the long-range $L^{-1}$ term.
Indeed Castleton \textit{et al.} noticed that finite-size scaling with potential alignment resulted in wide error bars. \cite{Castleton2006}
A series of simple cubic supercells containing 64, 216, 512 and 1000 atoms is chosen in the present study. 
The exceedingly large 1000-atom supercell restricts the calculations to the GGA-PBE functional, although we show that the obtained trend is applicable to hybrid functionals as well.
For the 64- and 216-atom cells, the Brillouin zone is sampled with a $2\times2\times2$ \textbf{k}-point mesh.
For larger supercells, we use two special \textbf{k}-points, \textit{i.e.} $\Gamma$-point (0,0,0) and $R$-point (0.5,0.5,0.5) in reciprocal coordinates.
Moreover, in the FNV scheme, the point charge $q_d$ consists of an exponential decaying term and a localized contribution modeled by a Gaussian 
\begin{equation}
\label{eq:fnv}
q_d(r) = q x N_\gamma e^{-r/{\gamma}} +q (1-x) N_\beta e^{-{r^2}/{\beta^2}},
\end{equation}
where $N_\gamma$ and $N_\beta$ are normalization constants, and $x$ is the fraction of the relative amount of the exponential decay.
In practice, the resulting corrected energy is insensitive to the choice of the specific parameters in Eq.~\eqref{eq:fnv}. \cite{Freysoldt2009} 

\begin{figure}
\includegraphics{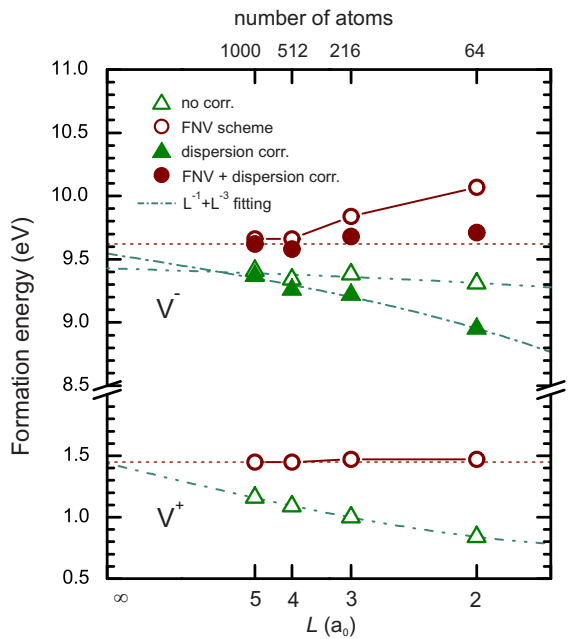}
\caption{\label{fig:corr}(color online). Demonstration of the correction schemes for the formation energies of the Cl vacancy in the $+1$ and $-1$ charge states (in Cl-rich limit) with respect to the reciprocal supercell lattice constant $L^{-1}$. The calculations were performed with the GGA-PBE functional without structural relaxations.}
\end{figure}

In Fig.~\ref{fig:corr}, we demonstrate the effects of finite-size corrections to the formation energies of Cl vacancies in +1 ($V^+$) and -1 ($V^-$) charge states.
No relaxation is taken into account at this stage so as to exclude the finite-size effect of elastic energies.
For $V^+$, one first notices that the extrapolated formation energy from the Makov-Payne scaling law falls in line with that of the FNV scheme.
The $L^{-1}$ clearly dominates for the Makov-Payne fitted curve.
The FNV scheme, on the other hand, shows a rapid convergence of the $V^+$ formation energy.
We see that finite-size correction is indeed mandatory for an accurate description of formation energy of charged defects. 
For the smallest 64-atom cell, the uncorrected formation energy is underestimated by roughly 0.6 eV.
Even for the 1000-atom cell, the formation energy without correction is still 0.2 eV too low.

\begin{figure}
\includegraphics{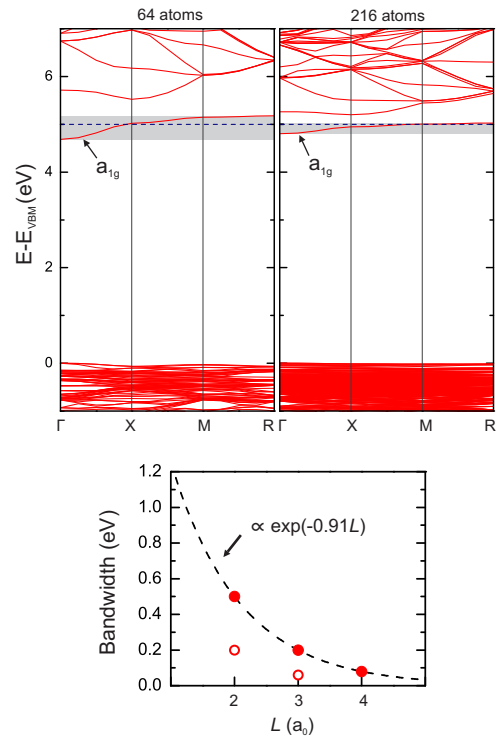}
\caption{\label{fig:band}(color online). Band structures (upper panel) of the Cl vacancy in the $-1$ charge state ($V^-$) in the unrelaxed 64- and 216-atom cells using the GGA-PBE functional. 
The shaded areas show the bandwidth of the doubly-occupied $a_{1g}$ state induced by the negatively charged defect.
The horizontal dashed line indicates the CBM of the perfect host without defects.
In the bottom panel, the evolution of the bandwidth of the $a_{1g}$ state (by closed circles $\bullet$) as a function of the supercell lattice constant is shown with an exponential fit curve for the unrelaxed structure.
The band dispersion for the relaxed cell is also given by open circles $\circ$.}
\end{figure}

Complexity arises when we move to the -1 charged Cl vacancy.
The uncorrected formation energies in Fig.~\ref{fig:corr} for $V^-$ exhibit a zigzag evolution with respect to the increasing supercell size, making the Makov-Payne fit unreliable.
Meanwhile, the FNV correction apparently yields a too high energy for small supercells, and the value does not appear to converge until we use the 512-atom cell.
The source of such error is identified as the spurious dispersion of the defect levels as a result of the overlap between the wavefunctions of the defect and its periodic images.
In the dilute limit, this localized defect level should be strictly a flat band.
However, as shown in Fig.~\ref{fig:band}, the Cl vacancy induced $a_{1g}$ level within the band gap shows a prominent dispersion for small supercells.
For hybrid functionals, the dispersion is less pronounced as the exact exchange favors a more localized electronic state in the vacancy. 
This artificial interaction tends to push the CBM of the host crystal to higher energies, or otherwise the CBM will become populated.
The defect level dispersion has a short-range characteristics, as is evidenced by the exponential fit of the $a_{1g}$ bandwidth with respect to the supercell lattice constant $L$ for the 64-, 216- and 512-atom cells (in Fig.~\ref{fig:band}).
This short-ranged effect is not included in either the Makov-Payne or FNV scheme, and thus one has to take it into account explicitly.
We note that the dispersion correction is not necessary for $V^+$ since its $a_{1g}$ level is unoccupied.
Here the correction for the dispersion is considered by calculating the energy difference between the KS energy of the $a_{1g}$ level at the $\Gamma$-point $\epsilon^\Gamma$ and the $a_{1g}$ KS energy averaged over the sampled \textbf{k}-points in the Brillouin zone $\overline{\epsilon}$.
The FNV scheme based on the dispersion corrected formation energies is again able to yield converged results for small supercells, and the results are also comparable to the Makov-Payne method including the dispersion effect in the dilute limit.

For the neutral Cl vacancy $V^0$, the situation becomes trouble-free since the electron is tightly bound to the vacancy center with a strongly localized electron density distribution as seen in Fig.~\ref{fig:chg}.
The formation energy barely varies for the supercells considered, and therefore there is no need for finite-size corrections for $V^0$.

We have shown the finite-size corrections for the charged defect supercells in rigid geometries with atoms fixed at their bulk positions.
However, the introduction of a Cl vacancy inevitably changes the electrostatic potential of the local environment, resulting in atomic relaxations around the vacancy.
The supercell approach, in this aspect, will lead to another error because the supercell employed in practice is usually not large enough for all local relaxations around the defect.
This error can be partially alleviated by restricting the relaxations to the first two atomic shells around the defect, although it might underestimate the relaxation energy.
Here we assess the finite-size effect on the elastic energy of the Cl vacancy in various charge states based on full relaxations using the GGA-PBE functional.
We do not discuss the relaxations of the outer shell atoms since their displacements are much smaller than the first shell Na atoms and they contribute little to the formation energy. 
For the neutral vacancy $V^0$, negligible inward relaxations of the six nearest neighbor Na atoms are found with supercells containing up to 216 atoms (see Table~\ref{disp}).
For $V^-$, the net negative potential induced by the excess electron added to the vacancy gives rise to an inward displacement of the neighboring cations.
As discussed in Sec.~\ref{sec:ks}, this results in a more localized $a_{1g}$ state, which consequently suppresses the dispersion of the $a_{1g}$ with respect to the rigid structure (see Fig.~\ref{fig:band}). 
Due to the finite-size of the supercell, Table~\ref{disp} shows that the displacement obtained from the 64-atom cell is 0.05 {\AA} smaller than that from the 216-atom cell.
The inability to fully relax in the 64-atom cell consequently gives a formation energy about 0.2 eV higher than that of the larger supercells.
For the positively charged vacancy $V^+$, the six nearest neighbor Na atoms experience an outward displacement due to the positive potential in the vacancy center.
In contrast to $V^-$, the finite-size effect on the elastic energy is not significant for $V^+$, as seen in Table~\ref{disp}, since the atomic displacements using the 64- and 216-atom cells are of similar magnitude.

With all these comprehensive finite-size effects in mind, we now summarize the correction scheme applied in the present work.
We restrict the calculations of formation energies with hybrid functionals to the use of the 64-atom supercell.
Thanks to the localized nature of the defect state and negligible relaxation for $V^0$, no correction is necessary.
For $V^+$ we apply the FNV to the 64-atom cell and refrain from any correction for the elastic energy.
For $V^-$ the dispersion correction is applied to the 64-atom cell, followed by the FNV scheme.  
Further, the relaxation energies are aligned with those obtained from the 216-atom cell, provided the latter already yields a converged elastic energy.
In practice, due to the similar amount of atomic displacement (see Table~\ref{disp}), we use the PBE result as a reference for the hybrid functional, and subsequently lower the formation energies by 0.2 eV for the relaxed 64-atom supercells in the -1 charge state.

\subsection{\label{sec:thermo}Chlorine vacancy thermodynamic transition energies}

\begin{table*}
\caption{\label{ef} Formation energies (in eV) of Cl vacancies in various charge states calculated with the GGA-PBE, HSE06 and mPBE0 functionals under Cl-rich and Cl-poor conditions. The Fermi energy is chosen at the VBM for charged defects. All values are corrected for finite-size effect.}
\begin{ruledtabular}
\begin{tabular}{ldddddd}
 & \multicolumn{3}{c}{Cl-rich conditions} & \multicolumn{3}{c}{Cl-poor conditions} \\
\cline{2-4}
\cline{5-7}
 & \multicolumn{1}{c}{GGA-PBE} & \multicolumn{1}{c}{HSE06} & \multicolumn{1}{c}{mPBE0} & \multicolumn{1}{c}{GGA-PBE} & \multicolumn{1}{c}{HSE06} & \multicolumn{1}{c}{mPBE0} \\
\hline
$V^0$ \\
rigid & 4.44 & 4.63 & 4.71 & 0.75 & 0.78 & 0.78 \\ 
relaxed & 4.44 & 4.63 & 4.71 & 0.75 & 0.78 & 0.78 \\
$V^+$ \\
rigid & 1.47 & 0.75 & -0.26 & -2.22 & -3.10 & -4.19  \\
relaxed & 0.63 & -0.10 & -1.10 & -3.05 & -3.95 & -5.03\\
$V^-$ \\
rigid & 9.71 & 11.40 & 12.91 & 6.02 & 7.55 & 8.98 \\
relaxed & 9.55 & 11.01 & 12.50 & 5.86 & 7.16 & 8.57\\
\end{tabular}
\end{ruledtabular}
\end{table*}

The calculated formation energies with finite-size corrections for the Cl vacancy in NaCl are shown in Table~\ref{ef}, with the Fermi energy $\epsilon_\text{F}$ fixed at the VBM.
For the neutral vacancy $V^0$, the hybrid functionals yield higher formation energies than the GGA-PBE, although the energy differences are small.
The functional dependence of formation energies for the charged states is much more prominent.
With the hybrid functionals, we obtain higher formation energies for $V^-$ and lower formation energies for $V^+$.
In particular, the negative formation energy of $V^+$ under Na-rich conditions suggests that the $F^+$ center could be predominating when the Fermi energy is close to the VBM.

To trace the source of the functional dependence of formation energy for charged vacancies, we may first rewrite the formation energy of $V^+$ in a rigid geometry with $\epsilon_\text{F}$ fixed at the VBM as
\begin{equation}
E_f(V^+) = (E_\text{D}^+ - E_\text{D}^0)+(E_\text{D}^0 - E_\text{H}^0 + \mu_\text{Cl}) + \epsilon_\text{VBM}, 
\end{equation}
where the first term on the right-hand side is the electron ionization energy of $V^0$ (or equivalently the affinity energy of $V^+$), and the second term simply the formation energy of $V^0$.
By decomposing the formation energy into several contributions, it is clear that the discrepancies in $E_f(V^+)$ stem mostly from the different positions of the VBM by various functionals.
We note that the ionization energy of $V^0$ shows very small changes (within 0.05 eV) from semi-local to hybrid functionals, consistent with the similar absolute energy of the singly occupied $a_{1g}$ state (see Table~\ref{ks}). 
Analogously, the formation energy of $V^-$ can be rewritten as 
\begin{equation}
E_f(V^-) = (E_\text{D}^- - E_\text{D}^0)+(E_\text{D}^0 - E_\text{H}^0 + \mu_\text{Cl}) - \epsilon_\text{VBM},
\end{equation}
where the first term on the right-hand side is now the (negative) electron affinity energy of $V^0$ (or the ionization energy of $V^-$).
In contrast to the ionization energy, it is found that hybrid functionals tend to yield a smaller affinity energy of $V^0$ than semi-local functionals, and that the difference reaches up to 0.9 eV.  
Along with the $\epsilon_\text{VBM}$, they explain the variations in the formation energy observed in Table~\ref{ef}.

The atomic relaxation energy due to the polaronic electron or hole can be further extracted from Table~\ref{ef} as the difference between the rigid and relaxed structures.
It is not surprising that the relaxation energies given by various functionals are consistent provided the atomic displacements are similar with these functionals (see Table~\ref{disp}).
The relaxation energy for $V^+$ is about 0.8 eV, while it ranges from 0.2 to 0.4 eV for $V^-$.

\begin{figure}
\includegraphics{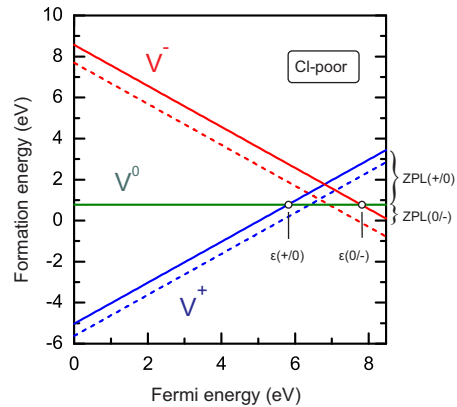}
\caption{\label{fig:ef}(color online). Calculated mPBE0 formation energies of the Cl vacancy with full relaxations as a function of the Fermi energy under the Cl-poor condition.
The solid lines represent the formation energies corrected for the finite-size effect. 
The thermodynamic transition levels and the zero-phonon lines (ZPL) are indicated.
The uncorrected values are given in dashed lines for reference.}
\end{figure}

For a charged defect, the formation energy is a function of the Fermi energy as illustrated in Fig.~\ref{fig:ef}.
The intersections of different charge states are the thermodynamic transition levels defined in Eq.~\eqref{eq:trans}.
We see in Fig.~\ref{fig:ef} that the transition levels $\epsilon$(+/0) and $\epsilon$(0/$-$) are both within the band gap.
Therefore, the mPBE0 functional predicts that all charge states (-1, 0 and +1) of Cl vacancy could be thermodynamically stable when the Fermi energy is varied within the band gap. 
We note that although this is also qualitatively predicted by the uncorrected formation energies, the Fermi energy window for the neutral $V^0$ vacancy is much narrower. 

\begin{figure}
\includegraphics{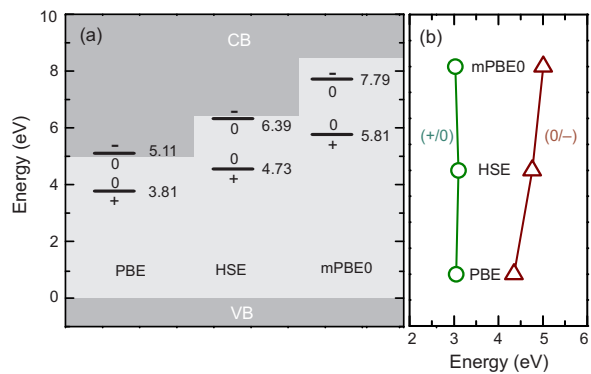}
\caption{\label{fig:trans} (color online). (a) The thermodynamic transition levels of the Cl vacancy calculated with various functionals.
The position of the VBM is aligned to energy zero.
(b) The absolute values of the thermodynamic transition energies.}
\end{figure}

The density functional dependence of the thermodynamic transition levels is illustrated in Fig.~\ref{fig:trans}(a).
Both neutral and +1 charge states are predicted to be stable since the $\epsilon$(+/0) levels are within the band gap for all functionals, with transition energies increasing from 3.81 eV to 5.36 eV as the band gap widens from semi-local to hybrid functionals.
On the other hand, the $\epsilon$(0/$-$) is placed slightly above the CBM in the PBE, while its position falls into the band gap in HSE06 and is further shifted downwards with respect to the CBM in the mPBE0 calculations.
Hence in contrast to the GGA-PBE, both the HSE06 and mPBE0 imply that the $V^-$ is stable.

While the thermodynamic transition energy generally increases with respect to the VBM as the band gap enlarges, we see from Fig.~\ref{fig:trans}(b) that the absolute position [$\epsilon(q/q')+\epsilon_\text{VBM}$] of the $\epsilon$(+/0) level remains roughly unaffected from semi-local to hybrid functionals.
This coincides with the findings by Alkauskas \textit{et al.} that the calculated energy levels of localized defect are generally not tied to the position of the CBM. \cite{Alkauskas2008}
The $\epsilon$(0/$-$) levels are nevertheless more dispersed.

\section{Optical properties of the color center}

\begin{figure}
\includegraphics{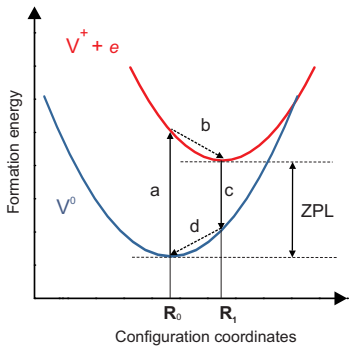}
\caption{\label{fig:cc}(color online). Configuration coordinate diagram for the neutral and $+$ charge state of Cl vacancy in NaCl. 
The Fermi energy is located at the CBM.
The optical processes involved are the absorption (a), emission (c), Stokes (b) and anti-Stokes (d) shifts, and zero-phonon line (ZPL).
The coordinates at the lowest vibrational state of the ground state and excited state are denoted by $\mathbf{R}_0$ and $\mathbf{R}_+$, respectively.
The zero-point energy is neglected in the diagram.}
\end{figure}

The experimentally available optical properties of the $F$ ($V^0)$ and $F'$ ($V^-$) center in NaCl serve as a benchmark for the assessment of the performance of the functionals.
The optical processes are clearly marked in the configuration coordinate diagram in Fig.~\ref{fig:cc} according to the Franck-Condon principle. \cite{Condon1928}
In the Franck-Condon approximation the electronic transition is assumed to occur very fast compared with the motion of nuclei in the lattice.
Therefore, the optical excitation spectrum observed in experiment does not involve the relaxation of the defect structure, in contrast to the thermodynamic transition.

The optical absorption and emission can be described by vibronic (simultaneous vibrational and electronic) transitions, in which the lattice vibration mode is treated by a quantum harmonic oscillator.
We first consider the excitation of an $F$ center, which is a well-defined feature. \cite{Pohl1937,Pick1938,Pick1958,Wood1964}
By absorption of a photon, the unpaired electron is transferred to an excited electronic state ($V^+$ state) and an excited vibrational state. 
The excitation of an electron into the CBM is equivalent to bringing an electron to a reservoir with a chemical potential of $\epsilon_\text{VBM}+E_g$.
The optical absorption energy $E_\text{a}$, as illustrated in Fig.~\ref{fig:cc}, is thus given by
\begin{equation}
\label{eq:abs}
E_\text{a} = E_f^{\mathbf{R}_0}(+;\epsilon_\text{F}=E_g) - E_f^{\mathbf{R}_0}(0),
\end{equation}
where the first term on the right-hand side is the formation energy of the unrelaxed +1 charge state (at coordinate $\mathbf{R}_0$) with the Fermi energy at the CBM, and the second term the formation energy of the relaxed neutral charge state (at $\mathbf{R}_0$).
The excited state ($F^+$ center) subsequently relaxes to its zero-point vibration states.
The energy gain due to the relaxation is the Stokes shift $E_\text{S}$ between the vertical absorption and the zero-phonon line (ZPL).
The ZPL is the transition energy from the lowest vibrational level (zero phonon mode) of the ground state to the lowest level of the excited state (at $\mathbf{R}_1$), without energy transfer to lattice phonons.
In terms of thermodynamic transition energy, it is easy to see from Fig.~\ref{fig:ef} that the ZPL can be expressed in terms of the difference between the band gap $E_g$ and $\epsilon$(+/0).
In the present case, due to the identical formation energy between the rigid and relaxed $V^0$, the Stokes shift reduces to the relaxation energy of the $V^+$ from the rigid structure.
In the vertical emission (luminescence), the excited electron from the CBM recombines into the defect level and the emission energy $E_\text{e}$ is given by
\begin{equation}
E_\text{e} = E_f^{\mathbf{R}_1}(+;\epsilon_\text{F}=E_g) - E_f^{\mathbf{R}_1}(0),
\end{equation}
where the defect structure in the neutral state is kept fixed as that in the relaxed +1 charge state.
Finite-size correction on this relaxation energy is taken into account according to Sec.~\ref{sec:corr}.
Once the electron is in the ground state, it relaxes to the bottom of the state with the relaxation energy $E_\text{AS}$ (\textit{i.e.} the anti-Stokes shift between the ZPL and the vertical emission).
The pronounced Stokes and anti-Stokes shifts (see Table~\ref{optics}) are expected due to the large polaronic distortion.

\begin{table}
\caption{\label{optics} Calculated vertical absorption ($E_\text{a}$) and emission ($E_\text{e}$) energies, zero-phonon line (ZPL) and the Stokes ($E_\text{S}$) and anti-Stokes ($E_\text{AS}$) shifts of the $F$ and $F'$ centers in NaCl.
All values in eV.}
\begin{ruledtabular}
\begin{tabular}{lddddd}
& \multicolumn{1}{c}{$E_\text{a}$} & \multicolumn{1}{c}{$E_\text{e}$} & \multicolumn{1}{c}{ZPL} & \multicolumn{1}{c}{$E_\text{S}$} & \multicolumn{1}{c}{$E_\text{AS}$} \\
\hline
$F$ center \\
GGA-PBE & 2.03 & 0.65 & 1.19 & 0.84 & 0.55 \\
HSE06 & 2.56 & 1.07 & 1.70 & 0.85 & 0.63 \\
mPBE0 & 3.50 & 1.88 & 2.66 & 0.84 & 0.78\\
Expt. & 2.77\footnotemark[1] & 0.98\footnotemark[2]& & & \\
\\
$F'$ center\\
GGA-PBE & 0.76 	& -0.27	& -0.11	& 0.87	& 0.16	 \\
HSE06 	& 0.91	& -0.34	& 0.04	& 0.86	& 0.38	 \\
mPBE0 	& 2.03	& 0.27	& 0.68	& 1.35	& 0.41	 \\
Expt. 	& 2.43\footnotemark[3] & & & & \\
\end{tabular}
\end{ruledtabular}
\footnotetext[1]{Reference \onlinecite{Bartram1968}.}
\footnotetext[2]{Reference \onlinecite{Ong1978}.}
\footnotetext[3]{Reference \onlinecite{La1966}.}
\end{table}

The calculated vertical absorption and emission energies of the $F$ center using the GGA-PBE and hybrid functionals are reported in Table~\ref{optics}.
The zero-point energy is not included since it is usually comparable for both the ground state and excited state.
It should be borne in mind that the total energy difference scheme ($\Delta$SCF) based on the one-particle picture is not capable of describing the electron-hole coupling in the optical absorption.
The presence of the exciton will usually introduce a uniform redshift of the absorption peak irrespective of the density functionals.

We find that the absorption and emission energies are underestimated by the GGA-PBE, and this is most likely related to the small band gap. 
At the present stage it remains unclear whether mPBE0 or HSE06 is more appropriate for the absorption energy since the excitonic binding energy of the $F$ center is yet unknown.
Meanwhile, the available experimental ZPLs for several $F$-aggregated centers (1.96 eV for $R_2$ band \cite{Fitchen1963} and 1.48 eV for $N$ band \cite{Pierce1964}) suggest the mPBE0 might give a too large ZPL.
Nevertheless, the hybrid functionals yield more realistic optical transition energies than the GGA functional.

Analogously, we extend the calculation to the optical process of the $F'$ center, which is formed when an electron is trapped at an $F$ center by light absorption at low temperatures. \cite{Pick1958}
Instead of the sharp and bell-shaped curve of the $F$ band, the $F'$ center of NaCl gives rise to a broad $F'$ absorption band, peaking at the longer wavelengths side of the $F$ band.\cite{Pohl1937,Pick1938,Pick1958}
It is seen in Table~\ref{optics} that all functionals now predict smaller absorption energies with respect to the experimental $F'$ band peak.
This tendency is not changed even if the excitonic effect is taken into account as the electron-hole interaction will further decrease the absorption energy.
In accord with the $F$ band absorption, the $F'$ $E_\text{a}$ increases from the semi-local functional to the hybrid functionals as the calculated band gap widens. 
The $E_\text{a}$ values given by the GGA-PBE and HSE06 are well below the experimentally observed peak, whereas the mPBE0 yields a value that is in better agreement with experiment. 
In addition, we find that the various functionals predict either a negative or a very small emission energy $E_\text{e}$ from the excited $F'$ state to the ground state.
A negative emission energy in Table~\ref{optics} suggests that the configuration coordinate of the intersection lies between the coordinates of the minimum of the ground state and excited state.
It is conceivable that in such case the excited state can return back to the ground state through a non-radiative process by vibrational relaxations, which leads to the luminescence quenching.
The non-radiative path is also valid for a vibronic system with a small emission energy where the intersection is in the vicinity of the minimum of the excited state.\cite{Georgiev1988}
Experimentally, a radiative transition of an $F'$ excited state is indeed absent.\cite{Benci1973}

\section{Discussion}
We have shown that, while hybrid functionals have been reported to be adequate for defects in some semiconductors,\cite{Saal2009,Stroppa2009,Gali2009,Agoston2009,Scanlon2009,Janotti2010,Deak2010} 
the description of the localized anion vacancy in a wide gap insulator is less satisfactory by hybrid functionals when compared to the experimental optical absorption spectra.
In this section, we aim to identify the possible origins of the failure of hybrid functionals for the description of the color centers in NaCl.

We start with the discussion of the optical absorption since it is well-defined experimentally.
The absorption energy of an $F$ center in Eq.~\eqref{eq:abs} can be rewritten as
\begin{eqnarray}
\label{eq:Ea}
E_\text{a} &=& -\epsilon(+/0)^{\mathbf{R}_0} + E_g \nonumber \\
           &=& \underbrace{[E_\text{D}^{\mathbf{R}_0}(+)-E_\text{D}^{\mathbf{R}_0}(0)]}_{\mbox{\text{IP}$(V^0)$}}+\underbrace{(\epsilon_\text{VBM}+E_g)}_{\mbox{$\epsilon_{\text{CBM}}$}},
\end{eqnarray}
where $\epsilon(+/0)^{\mathbf{R}_0}$ refers to the vertical transition energy occurring at the geometry for the neutral vacancy, and IP$(V^0)$ is the ionization energy of $V^0$.
Therefore the absorption energy is solely dependent on the ionization of the neutral vacancy and the position of the band edge in a perfect supercell, and no structural relaxation is involved.
As the ionization energy is not sensitive to the choice of the functional as discussed in Sec.~\ref{sec:thermo},
it becomes obvious that the discrepancies in the absorption energy reported in Table~\ref{optics} mostly stem from the variations in the CBM energy.
For instance, the GGA-PBE $\epsilon_\text{CBM}$ is 1.45 eV lower than the mPBE0 value as a result of the well-known band gap problem associated with the local and semi-local DFT functionals. 
While the energy gap can be reproduced by mixing 40\% exact exchange in mPBE0, it is yet not clear whether the positions of the band edges are accurate.

In principle, the band gap problem can be overcome by quasiparticle (QP) self-energy calculations based on many-body perturbation theory.
Here we follow the widely adopted $GW$ approximation for the electronic self-energy \cite{Hedin1965} and calculate the QP corrections to the Kohn-Sham eigenvalues.
The $GW$ approximation can be understood as the Hartree-Fock theory with a dynamically screened Coulomb interaction.
The QP energies are calculated in a two-atom NaCl unit cell with a $\Gamma$ centered $4\times4\times4$ $\mathbf{k}$-point mesh and an energy cutoff of 200 eV for the response function, and a total of 256 bands.
The dynamic dielectric matrix is constructed with a frequency grid of 200 points.\cite{Shishkin2006}
We note that the QP gap of NaCl is closely related to the starting wavefunction and self-consistency.
It is found that single shot $G_0W_0$ correction is too small when it is applied to the GGA-PBE eigenstates, whereas $G_0W_0$ on top of mPBE0 overestimates the QP gap.
A fully self-consistent $GW$ calculation also yields a too large QP gap irrespective of the initial eigenstates, as a result of the neglect of the attractive electron-hole interaction.\cite{Shishkin2007}
By updating the eigenvalues (four times) in the Green's function $G$ and keeping the screened Coulomb interaction $W$ at the RPA level within the initial PBE eigenvalues, the $GW_0$@PBE scheme produces a QP gap of 8.43 eV, in agreement with experiment.  
The VBM is now lowered by 2.77 eV with respect to the PBE eigenvalue, and the CBM is lifted up by 0.66 eV, reaching to 4.90 eV by QP corrections.
Compared to the $GW_0$ result, the mPBE0 CBM is placed 0.8 eV too high in energy, while the CBM energy calculated with the HSE06 functional coincides with that of the $GW_0$ (see Table~\ref{ks}).
A good agreement with the experiment $F$ band absorption energy can be already obtained if the $\epsilon_\text{CBM}$ in Eq.~\eqref{eq:Ea} is naively replaced by the $GW_0$ value while keeping the ionization energy untouched.
Therefore, the ionization energies of the neutral Cl vacancy $V^0$ calculated by semi-local and hybrid functionals are well described from the total energy difference method.
In contrast, the calculated ionization energy of the negative charge system $V^-$ is less satisfactory with GGA-PBE and is not much improved with the hybrid functionals based on the experimental $F'$ band absorption peak and the $GW_0$ CBM energy. 
This is easily understood since the electronic correlations for the removal of a second electron from the $a_{1g}$ level is beyond the scope of DFT.\cite{Bockstedte2010}
These many-body effects are accessible from the many-body perturbation theory, e.g. in the $GW$ approximation via the self-energy.

In the $GW$ approximation, the QP energies of the highest occupied and lowest unoccupied level correspond to the electron removal and addition energies, respectively.
It is then straightforward from Eq.~\eqref{eq:Ea} that the absorption energy can be calculated as the QP energy difference between the CBM and the lowest unoccupied state of the $V^+$, provided the ionization potential of the neutral defect system is equivalent to the electron affinity of the positive charged system. 
For example, using a 64-atom supercell with a cutoff energy of 100 eV and a $\Gamma$ point for the response function and 1024 total bands, the $GW_0$@PBE method yields an $F$ band absorption energy of 2.47 eV.
We note that the two-particle excitonic effect in the optical absorption is not taken into account in the GW approximation either.

For shallow defects, it has been found that the hybrid functional shows great improvement over local or semi-local functionals in the defect transition energies \cite{Deak2010}.
This is mostly likely benefited from the fact that the position of the shallow defect follows the band edge (either CBM or VBM), which can be reproduced by hybrid functionals with tunable $\alpha$.
For deep levels as demonstrated in this study, we find that a reproduction of a realistic band gap by an \textit{ad hoc} tuning of the amount of the exact exchange in hybrid functionals does not guarantee an accurate description of the optical defect levels, and the thermodynamic charge transition levels as well.
We see that the GGA-PBE is prone to an underestimation of the vertical transition energy, which is obviously impaired by a small band gap and a low conduction band edge.
A significant shift for the band edges can be observed with the hybrid functionals.
This leads to an increased vertical transition energy which is usually in better agreement with experiment, although the overestimation of the CBM energy by an increased fraction of the exact exchange in mPBE0 might give too high values (e.g. for the $F$ band).

In addition to the electronic contributions (e.g. ionization energy and the position of the band edge), the structural relaxation also plays an important role in predicting the transition energies.
The Frank-Condon shift (\textit{i.e.} the difference between the absorption and emission energy) sheds light on the effect of the exact exchange on the lattice relaxations around the vacancy.
By adopting the $GW_0$ CBM energy into the $F$ band emission energy in Table~\ref{optics}, we find that the lattice relaxation in presence of the electron-phonon interaction is best accounted by the mPBE0 hybrid functional as a result of the more localized electron density in the vacancy.
The localization is proportional to the amount of the non-local exact exchange which reduces the self-interaction arising from the DFT XC functional.
The localized nature of the trapped electron is further enhanced by a $GW$ calculation, exhibiting an even smaller $\mathbf{k}$-dispersion of the $a_{1g}$ level than the mPBE0 since the $GW$ approximation is free of self-interaction. 
In this context, we expect the structural relaxation will also be more realistic within many-body perturbation theory.

The discussion of the vertical transition energies finally invokes us to return to the thermodynamic transition level presented in Sec.~\ref{sec:thermo}.
The thermodynamic transition level can be readily decomposed into the electronic and structural contributions as 
\begin{equation}
\label{eq:epsilon_1}
\epsilon(+/0) = [E_g -E_\text{a}(F)] + \Delta E_\text{D}^+
\end{equation}
and
\begin{equation}
\label{eq:epsilon_2}
\epsilon(0/-) = [E_g -E_\text{a}(F')] + \Delta E_\text{D}^0
\end{equation}
where $\Delta E_\text{D}^+$ and $\Delta E_\text{D}^0$ are the relaxation energies from the initial geometry for the positive and neutral charge state, respectively.
By incorporating the experimental vertical absorption energy and mPBE0 relaxation energy into Eq.~\eqref{eq:epsilon_1} and \eqref{eq:epsilon_2}, 
we come to the thermodynamic transition energies $\epsilon$(+/0)=6.6 eV and $\epsilon$(0/$-$)=7.5 eV.
Therefore, the general picture of the Cl vacancy energetics by the adjusted hybrid functional so as to reproduce the experimental band gap remains qualitatively sound albeit not numerically accurate.

\section{Summary}
In this work we revisit the color centers in NaCl using hybrid density functionals.
The reduced self-interaction error alleviated by the exact exchange in hybrid functionals allows for a more accurate description of the atomic relaxations in the vicinity of the vacancy compared to the DFT results.
Yet, hybrid functionals are unable to achieve quantitative agreement with experimental optical absorption peaks.
We show that this is closely related to the overestimations of the band edges when the fundamental gap is reproduced by an empirical amount of exact exchange.
Meanwhile, when the electronic correlation comes into play during the removal (addition) of a second electron from (to) the localized defect level, the ionization (affinity) energy predicted within the framework of DFT is far from satisfactory.
More elaborate methods (such as $GW$ approximation\cite{Surh1995,Rinke2009,Martin-Samos2010,Lany2010} and two-particle Bethe-Salpeter equation\cite{Onida2002,Ma2008,Ma2010,Bockstedte2010}) are thus necessary for more accurate descriptions of the optical process and defect energetics of the localized defects in wide gap insulators.

\begin{acknowledgements}
We thank the allocation of computational time at H\"ochstleistungsrechenzentrum Nord (HLRN).
Financial support from K+S AG is gratefully appreciated.
\end{acknowledgements}

%

\end{document}